\documentclass{emulateapj}
\slugcomment{Accepted for Publication in Astrophysical Journal}
\lefthead{Rau et al.}
\righthead{A Survey for Fast Transients in the Fornax Cluster of Galaxies}

\def\asec{\ifmmode ^{\prime\prime}\else$^{\prime\prime}$\fi}
\def\farcs{\hbox{$.\!\!^{\prime\prime}$}}  	
\def\farcm{\hbox{$.\!\!^{\prime}$}}  
\def\degs{\ifmmode ^{\circ}\else$^{\circ}$\fi}
\def\farcm{\hbox{$^{\prime}$}}  

\shorttitle{Fornax Fast Transient Survey}
\shortauthors{Rau et al.}

\begin{document}


\title{A Survey for Fast Transients in the Fornax Cluster of Galaxies}


\author{A. Rau\altaffilmark{1}, E.O. Ofek\altaffilmark{1},   S.R. Kulkarni\altaffilmark{1}, B.F. Madore\altaffilmark{2}, O. Pevunova\altaffilmark{3}, M. Ajello\altaffilmark{4}}

\email{arne@astro.caltech.edu}

\altaffiltext{1}{Caltech Optical Observatories, MS 105-24, California Institute of Technology, Pasadena, CA 91125, USA}
\altaffiltext{2}{Observatories of the Carnegie Institution of Washington, 813 Santa Barbara Street, Pasadena, CA 91101, USA}
\altaffiltext{3}{Infrared Processing and Analysis Center, MS 100-22, California Institute of Technology, Jet Propulsion Laboratory, Pasadena, CA 91125, USA}
\altaffiltext{4}{Max-Planck Institut f\"ur Extraterrestrische Physik, Giessenbachstr.1, 80748 Garching, Germany}

\begin{abstract}

  The luminosity gap between novae ($M_{\rm R}\le -10$) and supernovae
  ($M_{\rm R}\ge -14$) is well  known since the pioneering research of
  Zwicky and Hubble.  Nearby  galaxy clusters and concentrations offer
  an  excellent opportunity  to  search for  explosions brighter  than
  classical novae  and fainter than  supernovae. Here, we  present the
  results of  a $B$-band  survey of 23  member galaxies of  the Fornax
  cluster, performed at the Las Campanas 2.5-m Irene duPont telescope.
  Observations  with a  cadence of  32\,minutes discovered  no genuine
  fast  transient   to  a  limiting  absolute   magnitude  of  $M_{\rm
    B}=-9.3$\,mag.  We provide a  detailed assessment of the transient
  detection  efficiency and the  resulting upper  limits on  the event
  rate  as  function  of  peak  magnitude.  Further,  we  discuss  the
  discoveries of five previously unknown foreground variables which we
  identified as two flare stars, two W~Uma type eclipsing binaries and
  a candidate $\delta$~Scuti/SX~Phe star.

\end{abstract}



\section{Introduction}

The recent  findings of  a number of  enigmatic transients,  e.g., new
types of  novae \citep{Kulkarni:2007uq} and  supernovae related events
\citep{Ofek:2007fj,Smith:2007qy,Quimby:2007qy,Pastorello:2007jk}, have
demonstrated that  the phase space  of eruptive transients  is already
richer  than discussed  in  astronomy texts.   Even more  discoveries,
especially on  timescales of  days to weeks,  are anticipated  for the
upcoming     large-area      facilities,     such     as     SkyMapper
\citep{Schmidt:2005qy},  the  Panoramic  Survey  Telescope  and  Rapid
Response  System \citep[Pan-STARRS;][]{Kaiser:2002fk},  and  the Large
Synoptic Survey  Telescope \citep[LSST;][]{Tyson:2005lr}.  However, on
shorter timescales (minutes to hours), their observational designs are
less optimal and dedicates, small  experiments are more likely to lead
to great findings.  A small number of such surveys have been performed
previously
\citep[e.g.,][]{Becker:2004fk,Rykoff:2005lr,Morales-Rueda:2006qy,Ramsay:2006lr}
and informed us  of the difficulty of discovering  genuine new classes
of fast transients and variables\footnote{In the following we refer to
  an event as transient when  it has no known quiescent counterpart at
  any wavelength and  to a source which shows  a quiescent counterpart
  in our images or has  a catalogued counterpart as variable.}.  Here,
the  greatest challenge  is  to  pierce the  fog  of known  foreground
contaminators.   Asteroids   and  flares  from  M   dwarfs  have  been
established   to    dominate   the   short-timescale    variable   sky
\citep{Kulkarni:2006lr}.   However, once  this curtain  is penetrated,
exciting  and  rare events  can  be  found,  e.g., the  high-amplitude
optical  flickering   ($\sim$3.5\,mag  in  6\,min)   accompanying  the
outburst   of  the   Galactic   X-ray  binary   Swift~J195509.6+261406
\citep{Stefanescu:2007fk,Kasliwal:2007qy}.   The   latter  source  was
brought    to   attention   by    its   preceding    gamma-ray   flare
\citep{Pagani:2007fk}.  Nonetheless, high-cadence optical surveys will
be  capable   of  detecting   similar  events  independently   of  the
high-energy emission.

Another,  yet  unidentified,  bright  ($R=11.7$\,mag),  high-amplitude
candidate  transient  ($>$6\,mag in  2\,min)  was recently  discovered
along the line  of sight to the nearby  ($\sim$70\,Mpc) galaxy IC~4779
\citep{Klotz:2007lr}. If associated with  IC~4779, the event reached a
staggering peak  luminosity of $L_{\rm R,peak}\sim3\times10^{44}$\,erg
s$^{-1}$,   outshined   only   by   the  brightest   known   supernova
\citep[SN2005ap;][]{Quimby:2007qy}.

A  critical  component  of  every  successful transient  search  is  a
judicious   target  selection.   This   is  especially   relevant  for
short-cadence  experiment, where  only  a small  number  of fields  is
repeatedly  observed.  Obvious targets  for such  a search  are nearby
massive clusters of galaxies.   Clusters contain a large stellar mass,
which  increases the  probability of  finding rare  events  during the
transient phase.  Furthermore, their  known distance provides a direct
access  to the  absolute brightness  of the  events. In  addition, the
variety  of galaxy  types,  and thus  stellar  populations, may  offer
constraints on the progenitor ages.

In this paper we present the  results of an optical survey of galaxies
in the Fornax Cluster designed  to test the transient and variable sky
in a galaxy  cluster environment on timescales shorter  than one hour.
At    the    distance    of    $16.2\pm1.5$\,Mpc    \citep[$(m-M)_{\rm
  0}=30.91\pm0.19$;][]{Grillmair:1999lr}     and    with    negligible
foreground extinction  \citep[E(B-V) = 0.013\,mag;][]{Schlegel:1998ul}
observations  of the  Fornax cluster  allow one  to probe  the  gap in
absolute peak brightness between classical novae (M$_{\rm R}>-10$) and
supernovae (M$_{\rm  R}<-14$) already with  2-m-class telescopes. This
part  of  phase  space  has  become  particularly  alluring  with  the
recognition  of an  emerging population  of new  types  of explosions,
namely           the            Luminous           Red           Novae
\citep{Kulkarni:2007uq,Rau:2007kx,Ofek:2007uq}.


\section{Observations and Data reduction}
\label{sec:data}

Observations were  obtained with the  Wide Field Reimaging  CCD Camera
(WFCCD) at the  2.5-m Irene du Pont telescope  in Las Campanas, Chile.
The detector dimensions of  2048$\times$2048 pixels, together with the
plate  scale of  0\farcs774, provide  a  circular field  of view  with
$\sim$12\farcm5 radius.   Imaging was performed  in the $B$-band  as a
compromise between avoiding  flares from Galactic M-dwarfs (especially
bright  in  the  ultra-violet)  and increasing  the  contrast  between
possible transients and the  predominantly early type cluster galaxies
(bright in the red part of the spectrum).

\begin{figure}[htbp]
\begin{center}
\includegraphics[width=0.35\textwidth,angle=-90]{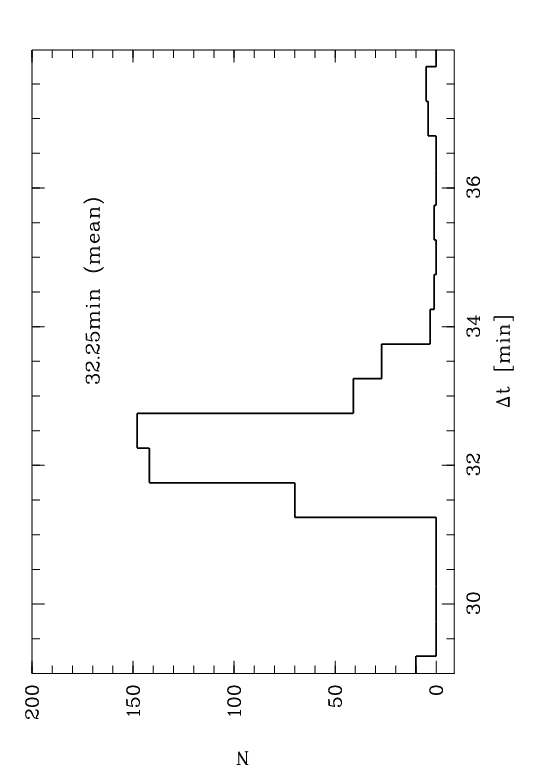}
\caption{Time between two consecutive images of a given pointing. The mean cadence is 32.25\,min.}
\label{fig:cadence}
\end{center}
\end{figure}

For  the survey,  we selected  ten  fields covering  the 23  brightest
Fornax~I        cluster        members        (Table~\ref{tab:fields},
Figures available in printed paper).    These   fields   were
imaged  with a  sequence of  120\,s exposures  repeated up  to fifteen
times in each of five nights in October and December 2006.  Additional
single exposures  were obtained in  five nights in November  2006 (see
Table~\ref{tab:obslog} for a  complete log). As part of  a sequence, a
single exposure was taken at  each position, after which the telescope
slewed to the next field.  The resulting distribution of times between
consecutive     images    for     each    field     is     given    in
Figure~\ref{fig:cadence}.   Throughout  the  survey a  mean  intra-day
cadence  of $\Delta$$t=32.25$\,min  was achieved.   The spread  of the
observations  between  October  and  December  2006  allowed  for  the
possible detection of slowly  evolving events with timescales of about
one month (e.g., supernovae).

\begin{table}
\begin{center}
\caption{Fields.
\label{tab:fields}}
\begin{tabular}{clll}
\tableline\tableline
\# & RA$_{\rm 2000\tablenotemark{a}}$ & Dec$_{\rm 2000}\tablenotemark{a}$ & Galaxies\tablenotemark{b} \\
\tableline
A & 03:24:36.7 & -36:26:21&  NGC1326, 1326A, 1326B\\
B & 03:22:41.1 & -37:10:54 & NGC1316, 1317\\
C & 03:31:04.3 & -33:36:42 & NGC1350\\
D & 03:36:29.5 & -34:52:28 & NGC1380, 1380A\\
E & 03:35:44.1 & -35:18:50 & NGC1373, 1374, 1375, 1379, 1381\\
F & 03:36:49.0 & -35:20:55 & NGC1379, 1381, 1382, 1387, MCG-06-09-008\\
G & 03:38:34.0 & -35:29:58 & NGC1399, 1404\\
H & 03:36:45.2 & -36:06:17 & NGC1369, 1386\\
I & 03:33:37.5 & -36:07:19 & NGC1365\\
K & 03:42:21.9 & -35:16:12 &NGC1427, 1428\\
\tableline
\end{tabular}
\tablenotetext{a}{Coordinates of the centers of the fields.}
\tablenotetext{b}{Prominent galaxies within the field of view.}
\end{center}
\end{table}


\begin{table}
\begin{center}
\caption{Observation log.
\label{tab:obslog}}
\begin{tabular}{ccccccccccc}
\tableline\tableline
Date (UT)\tablenotemark{a} & \multicolumn{10}{c}{Exposures\tablenotemark{b}}\\ 
		& A	&B	&C	&D	&E	&F	&G	&H	&I	&K\\
\tableline
06/10/23	&10	&10	&10	&10	&10	&9	&8	&8	&8	&10\\
06/10/24	&13	&14	&13	&12	&14	&10	&13	&13	&13	&13\\
06/10/25   &\multicolumn{10}{c}{lost due to bad weather}\\
06/11/17	&1	&1	&1	&1	&1	&1	&1	&1	&1	&1\\
06/11/18	&1	&1	&1	&1	&1	&1	&1	&1	&1	&1\\
06/11/19	&1	&1	&1	&1	&1	&1	&1	&1	&1	&1\\
06/11/20	&1	&1	&1	&1	&1	&1	&1	&1	&1	&1\\
06/11/21	&1	&1	&1	&1	&1	&1	&1	&1	&1	&1\\
06/12/19	&14	&15	&14	&14	&14	&14	&14	&14	&14	&14\\       
06/12/20	&13	&13	&13	&14	&13	&14	&14	&14	&14	&14\\              
06/12/21   &12	&12	&12	&12	&12	&12	&12	&10	&12	&12\\
\tableline
total\tablenotemark{c}: 	&67	&69	&67	&67	&68	&64	&66	&64	&66	&68\\
\tableline
\end{tabular}
\tablenotetext{a}{UTC date of observation in the format YY/MM/DD.}
\tablenotetext{b}{Number of 120\,s $B$-band exposure for each field in a given night.}
\tablenotetext{c}{Total number of observations for each field.}
\end{center}
\end{table}

Figure~\ref{fig:cadence} indicates that  our primary sensitivity is to
variability  on  $\sim32$\,min  timescales.   Here, we  have  a  total
exposure time of 13.7 days distributed over 616 images (neglecting the
once-per-day  exposures   in  November).   With  a  covered   area  of
0.136\,deg$^2$ per  image, this yields a total  areal exposure, E$_A$,
of 1.86\,deg$^2$ days.  We have a  factor of 5 and 44 larger E$_A$ for
inter-night and inter-month observation intervals, respectively.

Image reduction (bias subtraction,  flat fielding) was performed using
standard  {\it  IRAF}  routines\footnote{IRAF  was distributed  by  the
  National Optical Astronomy Observatories,  which are operated by the
  Association of  Universities for Research in  Astronomy, Inc., under
  cooperative  agreement with the  National Science  Foundation}.  The
astrometric   solutions   were   obtained   in  reference   to   NOMAD
\citep{Zacharias:2005lr}       with       {\it      ASCfit       v3.0}
\footnote{http://www.astro.caltech.edu/$\sim$pick/ASCFIT/README.ascfit3.html}.
The  strategy  to  search  for  transients  and  variables  in  single
exposures  implies  that Cosmic-ray  removal  by  median combining  of
multiple  images was not  applicable.  Hence,  we applied  a Laplacian
algorithm  \citep[{\it lacos\_im};][]{van-Dokkum:2001qy}  which allows
the identification of Cosmic-rays in single frames.

\begin{figure}[htbp]
\begin{center}
\includegraphics[width=0.45\textwidth,angle=0]{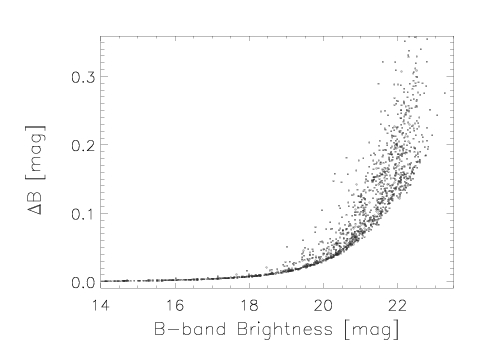}
\caption{Photometric accuracy. Shown are the combined point source measurements of three exposures obtained on 2006 Dec 19 UT. The seeing was 1\farcs4 and a 5-$\sigma$ limiting magnitude of $B\sim$22\,mag was reached. Sources brighter than $B\sim15$\,mag are saturated.}
\label{fig:deltaMag}
\end{center}
\end{figure}

For  absolute  photometric calibration  we  used  observations of  the
standard   star  field   T~Phe   \citep{Landolt:1992lr}  taken   under
photometric  conditions  (2006  November  19).  This  calibration  was
applied to  the images  of the ten  Fornax fields obtained  during the
same night.  Exposures from  the remaining nights were tied relatively
to these absolutely calibrated  frames.  Here, we selected an ensemble
of   at   least  40   local,   non-saturated,  non-variable   ($\Delta
B<0.01$\,mag) reference stars for  each image.  The photometric offset
between  the two  observations was  estimated by  deriving  the median
brightness  offset  of these  sets  of  stars  with respect  to  their
brightness in the reference frame.  This was successively done for all
observations  of a given  field, thus  providing a  common photometric
zero-point.  The  resulting photometric accuracy for  point sources is
shown in  Figure~\ref{fig:deltaMag}.  In a typical  night a 5-$\sigma$
limiting magnitude  of $B~\sim22$\,mag was  reached.  Sources brighter
than $B\sim15$\,mag were generally saturated.

The search for transients and  variables was performed in two separate
steps. First  we applied a  point spread function  (PSF)-matched image
subtraction   using   a  software   package   based   on  {\it   ISIS}
\citep{Alard:2000fk}.  Reference frames  for all fields were generated
by combining the three best  quality exposures taken in November 2006.
Candidate  transients and  variables were  detected in  the difference
images  using  {\it SExtractor  v2.5.0}  \citep{Bertin:1996uq}. For  a
typical  image,  this resulted  in  about  100  sources brighter  than
5-$\sigma$, out of  which all but a few  (see below) were subsequently
identified  by   eye  as  false  positives   (e.g.,  image  artifacts,
Cosmic-ray residuals, PSF distortion, and  sources near the rim of the
circular  field  of view  which  were  in the  input  but  not in  the
reference  frame).  The  PSFs of  the resulting  candidates  were then
verified against those of stars in the input images.

\begin{figure}[htbp]
\begin{center}
\includegraphics[width=0.50\textwidth,angle=0]{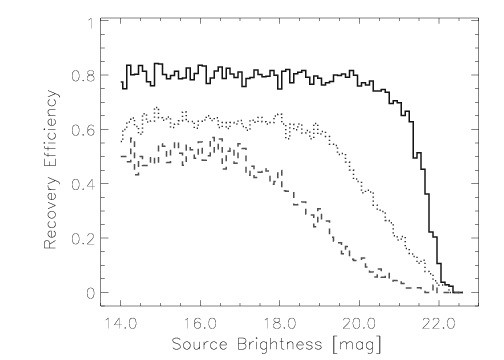}
\hspace{-0.9truecm}
\includegraphics[width=0.50\textwidth,angle=0]{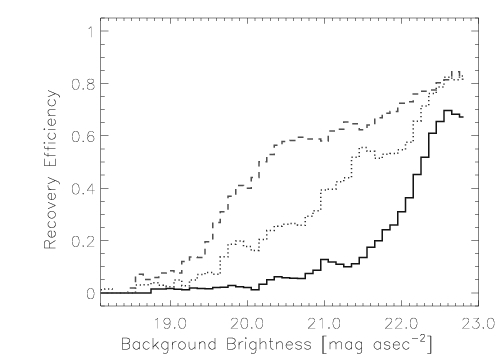}
\caption{left: Transient recovery efficiency as function of source magnitude for three levels of background surface brightness. The solid line (B=22.4\,mag asec$^{-2}$) corresponds to the efficiency away  from the Fornax cluster members, while the dotted (B=21.5\,mag asec$^{-2}$) and dashed (B=20.0\,mag asec$^{-2}$) lines are representative for locations within the galaxies. Sources brighter than B=15 are generally saturated. The faint sources cutoff  corresponds to the 5-$\sigma$ detection limits in the difference images. 
right: Transient recovery efficiency as function of background surface brightness for test sources with B=21.3 (solid), 19.5 (dotted) and 18.0 (dashed). Fewer efficiency points have been obtained at high surface brightness regions, thus, these bins have the lowest statistic.}
\label{fig:efficiency}
\end{center}
\end{figure}

The efficiency, $\epsilon$, of recovering transients in the difference
images was  modeled with Monte  Carlo simulations.  Here, a  subset of
the survey data was enriched with PSF-matched artificial point sources
and subsequently  passed through the image  subtraction pipeline.  The
efficiency was  then calculated from  the ratio of recovered  to input
test sources.  In total we obtained 1.9 Million efficiency points with
random location in the field and random magnitudes of 14$<$B$<$23.  As
the main  driver for our  survey was the  search for events  in nearby
galaxies, understanding  $\epsilon$ as  function of the  local surface
brightness became  important. Thus, a subset of  the efficiency points
was placed inside the extend of the bright Fornax cluster members.  As
expected, $\epsilon$ was found  to decrease with increasing background
flux  contribution  (see  Fig.~\ref{fig:efficiency}a).  For  ``empty''
locations  ($B\sim22.6$\,mag  asec$^{-2}$)  a recovery  efficiency  of
$\epsilon=0.85$ (0.7)  was achieved  for sources brighter  than $B=20$
(21.3).   In positions  with higher  surface  brightness ($B\sim21.5$,
$20.0$\,mag   asec$^{-2}$)   the    maximum   efficiency   was   lower
($\epsilon=0.60$, $0.40$).  Fig~\ref{fig:efficiency}b shows $\epsilon$
as function of surface brightness for three test source magnitudes.

In   a  second   step   we  performed   a  catalog-based   variability
search. While this method is inferior to image subtraction for sources
overlaying the bright cluster galaxies, it improves the detectivity of
low-amplitude,  bright  variables  in  the field.   For  these,  image
subtraction is typically very sensitive to PSF stability, convolution,
and alignment,  and residuals  unrelated to intrinsic  variability may
remain.   The  source catalogs  were  compiled  using {\it  SExtractor
  v2.5.0}  on  the  photometrically  calibrated  images  (see  above).
Objects  with  two  or   more  detections  brighter  than  $B=20$  and
deviations of more than $B=0.1$\,mag from their median brightness were
selected as candidate variables.

A serious contamination in optical transient and variable searches can
come  from solar  system bodies.   However, due  to the  high ecliptic
latitude ($\beta\approx-53$\,\degs) observations of the Fornax cluster
are only  marginally affected. Our  choice of using the  B-band filter
further  reduced the  impact  of asteroids,  whose  emission peaks  at
longer   wavelength.   Moreover  repetitive   visits  of   each  field
throughout the nights  were allowing secure astrometric identification
of moving objects  with proper motions as low as  3\asec\ per day. For
example, the  expected parallax  of a Kuiper  belt object,  at 100\,AU
from the Sun  at that location and time of  the year, is $\sim33$\asec
day$^{-1}$ and $\sim3\farcs5$\,day$^{-1}$ due  to the Earth and object
motion,  respectively.  Thus, it  was  not  surprising  that only  one
astrometrically variable was found in the survey dataset.
 

\section{Results}

The  search netted  seven high-confidence  photometric  transients and
variables.  Two,  both  earlier  reported  type Ia  SNe  in  NGC~1316,
SN~2006dd  \citep{Monard:2006lr} and  SN~2006mr \citep{Monard:2006fk},
were only  detected in image subtraction.  The  remaining five objects
were variable  point sources found with both  detection methods.  None
of        them        was        previously       catalogued        in
SIMBAD\footnote{http://simbad.u-strasbg.fr/},
NED\footnote{http://nedwww.ipac.caltech.edu/},
GCVS\footnote{heasarc.gsfc.nasa.gov/W3Browse/all/gcvsnsvars.html}    or
detected in the  {\it ROSAT}/PSPC all-sky survey \citep{Voges:1999fj}.
Below  we provide  a  brief  description for  each  of these  variable
sources (see also Table~\ref{tab:candidates}).

\begin{table*}
\begin{center}
\caption{Summary of detected sources.
\label{tab:candidates}}
\begin{tabular}{lllcccl}
  \tableline\tableline
  \# & RA$_{\rm 2000}$ & Dec$_{\rm 2000}$ & B$_{\rm peak}$\tablenotemark{a} & B$_{\rm quiescence}$\tablenotemark{b} &Classification & D$_{\rm proj}$\tablenotemark{c} \\
  & & & (mag) & (mag) & & \\
  \tableline
  FA-1   &  03:24:24.56 & $-$36:30:26.5 & 18.91$\pm$0.03 & 19.17$\pm$0.03 & M1--2 & 36\,kpc to NGC~1326\\
  FB-1    & 03:22:31.38 & $-$37:04:01.5 &  18.48$\pm$0.02 & 18.84$\pm$0.14 & (?)$\delta$~Scuti/SX~Phe & 26\,kpc to NGC~1317\\ 
FE-1    & 03:36:18.85 & $-$35:14:58.9 & 16.07$\pm$0.01 & 16.18$\pm$0.09 & W~Uma & 29\,kpc to NGC~1381\\
FH-1    & 03:36:53.67 & $-$36:05:29.9 & 18.66$\pm$0.03 & 20.70$\pm$0.09 & M3--4 & 42\,kpc to NGC~1386\\
FK-1    & 03:42:33.73 & $-$35:18:29.0 & 15.20$\pm$0.01 & 15.42$\pm$0.20 & W~Uma & 35\,kpc to NGC~1427\\
SN~2006dd & 03:22:41.64 & $-$37:12:13.2 & 17.25$\pm$0.03\tablenotemark{d} & -- & SN Ia & 1.2\,kpc to NGC~1316 \\
SN~2006mr & 03:22:42.84 & $-$37:12:28.5 & 16.16$\pm$0.03\tablenotemark{e} & -- & SN Ia & 1.1\,kpc to NGC~1316 \\
\tableline
\end{tabular}
\tablenotetext{a}{Observed peak $B$-band magnitude.}
\tablenotetext{b}{Observed mean (for periodic variables) or quiescence (for eruptive variables) $B$-band magnitude.}
\tablenotetext{c}{Projected distance to center of nearest Fornax cluster galaxy.}
\tablenotetext{d}{Observed $B$-band magnitude on 2006 Oct 23.13 UT}
\tablenotetext{e}{Observed $B$-band magnitude on 2006 Nov 17.58 UT}
\end{center}
\end{table*}

\subsection{Eruptive variables}

\begin{figure*}[htbp]
\begin{center}
\includegraphics[width=1\textwidth,angle=0]{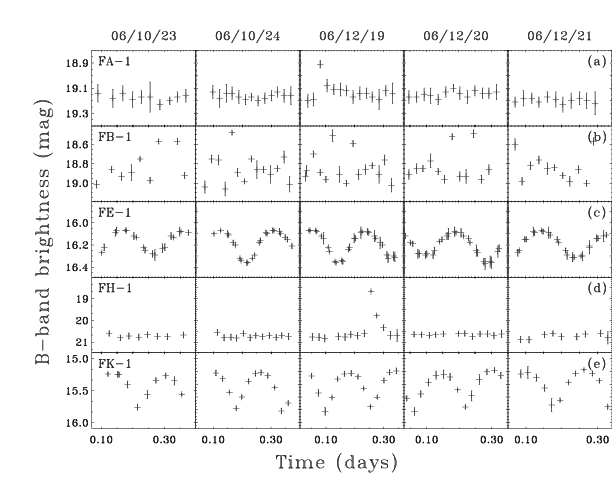}
\caption{Observed $B$-band measurements for the five candidate variables listed in Table~\ref{tab:candidates}. Shown are the lightcurves for the five nights with multiple exposures per field (2006 Oct 23, 24 and Dec 19, 20, 21 UTC). Time is given in fractional days (UTC). A compilation of all data is available in electronic form (Table~\ref{tab:data}).}
\label{fig:lc}
\end{center}
\end{figure*}

Two of the  variable sources were detected when  they exhibited single
flares from their otherwise  constant quiescent brightness. The first,
FA-1  (Figure~\ref{fig:fc}a),  is  a  faint,  $B=19.17\pm0.03$,  point
source    that    showed   a    distinct    outburst   with    $\Delta
B=0.26\pm0.04$\,mag (Figure~\ref{fig:lc}a). The  rise time to peak was
shorter  than 1\,hr  (2$\times$cadence)  and the  decline lasted  over
$\sim2.5$\,hrs.   The  source  has  a bright  near-IR  counterpart  as
detected     by     2MASS\footnote{http://www.ipac.caltech.edu/2mass/}
($J=14.58\pm0.03$,      $K=13.65\pm0.04$).      We      fitted     its
$B-J=4.59\pm0.04$\,mag            and           $B-K=5.52\pm0.05$\,mag
colors\footnote{Here,  we  assume  that  the  2MASS  observations  are
  representative  of  the  quiescence   (or  median  in  case  of  the
  non-eruptive  sources  discussed  in  \S~\ref{sec:per})  brightness.
  This  is supported  by  generally lower  amplitudes  in the  near-IR
  compared    to    the    $B-$band.}     with    stellar    templates
\citep{Pickles:1998xy} and  found them to be consistent  with an M1--2
star.  The most likely explanation is a low-amplitude long-decay flare
of a UV Ceti type Galactic M1--2 dwarf (M$_{\rm B}\sim11.5$\,mag) at a
distance of $\sim350$\,pc.

FH-1 (Figure~\ref{fig:fc}d)  was detected  when it underwent  a strong
flare with $\Delta$B=2.06$\pm$0.15\,mag.  It rose to peak in less than
1\,hr and decayed to its quiescent brightness of $B=20.70\pm0.09$ over
1.5\,hrs  (Figure~\ref{fig:lc}d).   FH-1  is  bright  in  the  near-IR
($J=15.02\pm0.02$,      $K=14.29\pm0.07$)      and     its      colors
($B-J=5.68\pm0.10$\,mag,  $B-K=6.41\pm0.10$\,mag)  indicate  an  M3--4
stellar classification. We suggest that this outburst was a long-decay
flare of  an M3--4-dwarf (M$_{\rm  B}\sim12.5$\,mag) at a  distance of
$\sim$450\,pc .

\subsection{Periodic variables}
\label{sec:per}

\begin{figure}[htbp]
\begin{center}
\includegraphics[width=0.35\textwidth,angle=90]{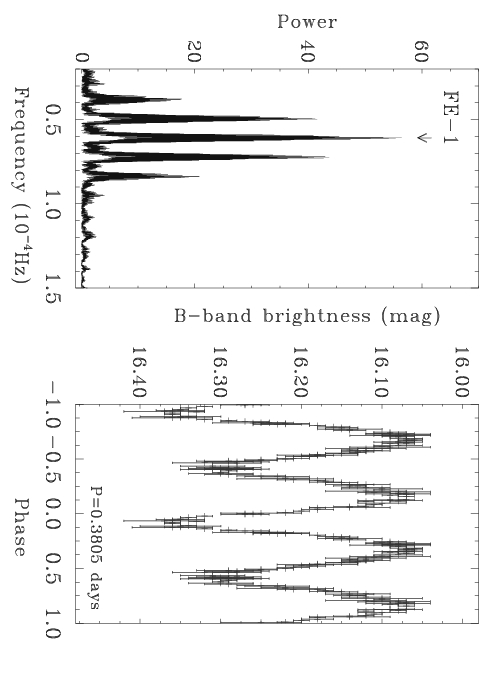}
\caption{Lomb-Scargle periodogram (left) for FE-1 showing that the highest power occurs for at a frequency of $(6.083\pm0.001)\times10^{-5}$\,Hz (or period of $0.19025\pm0.00002$\,days). The phase  folded  heliocentric  corrected lightcurve (right) indicates differences in the depth of consecutive minima, suggesting that the real period is likely at twice the above value, $0.38050\pm0.00004$\,days.} 
\label{fig:phaseE}
\end{center}
\end{figure}

\begin{figure}[htbp]
\begin{center}
\includegraphics[width=0.45\textwidth,angle=0]{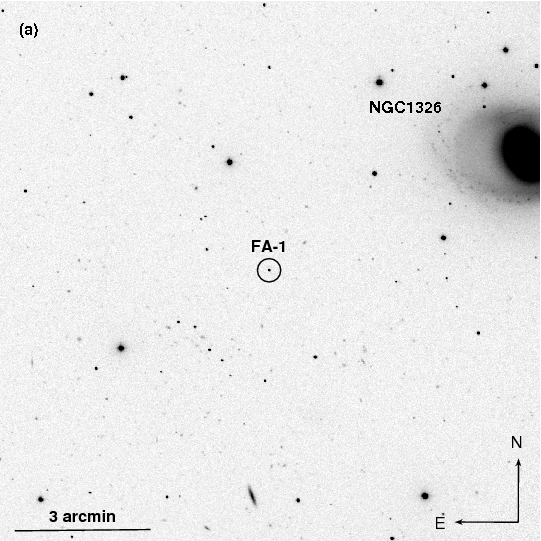}
\includegraphics[width=0.45\textwidth,angle=0]{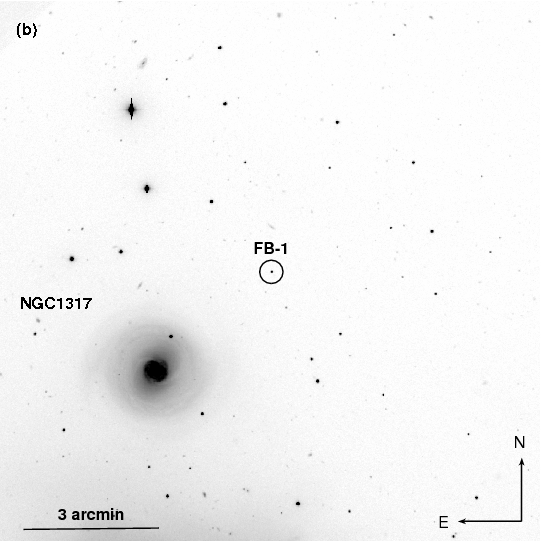}
\includegraphics[width=0.45\textwidth,angle=0]{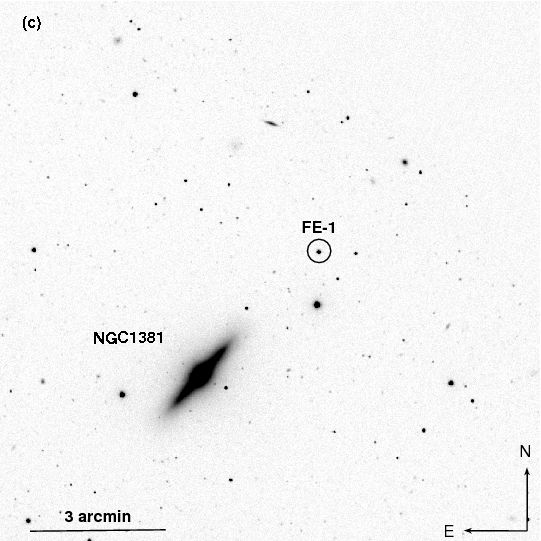}
\includegraphics[width=0.45\textwidth,angle=0]{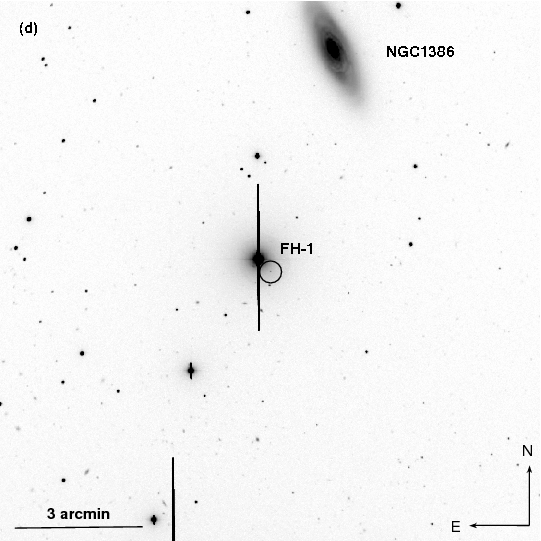}
\includegraphics[width=0.45\textwidth,angle=0]{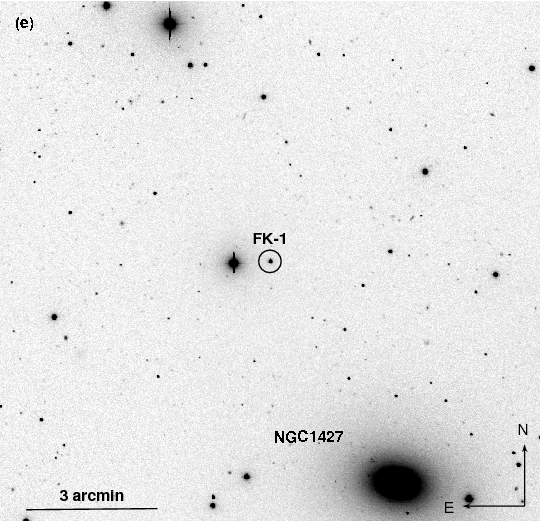}
\caption{Finding charts.}
\label{fig:fc}
\end{center}
\end{figure}

Regular  brightness modulations  were  shown by  three sources.   FE-1
(Figure~\ref{fig:fc}c)  displayed   a  nearly  sinusoidal  variability
(Figure~\ref{fig:lc}c)   with   a   maximum   amplitude   of   $\Delta
B=0.31\pm0.05$\,mag.          A        Lomb-Scargle        periodogram
\citep{Scargle:1982kx}   of  the  heliocentric   corrected  lightcurve
revealed  two possible periods,  one at  $0.19025\pm0.00002$\,days and
another    at    twice    this    value    ($0.38050\pm0.00004$\,days;
Figure~\ref{fig:phaseE}).   The  latter  is  favored  by  an  apparent
difference of $\Delta B=0.07\pm$0.02\,mag  in depth of two consecutive
minima.  Note  that FE-1  was included  in the overlap  of two  of our
pointings (FE \& FF) and thus has twice the data coverage (132 images)
compared to  the remaining candidates.   2MASS detected the  source at
$J=15.53\pm0.05$  and $K=15.16\pm0.15$,  which, together  with  a mean
brightness of $B=16.18\pm0.09$,  results in $B-J=0.65\pm0.10$\,mag and
$B-K=1.02\pm0.17$\,mag.   These  colors  suggest  a spectral  type  of
A7--F0.     The    source   is    also    included    in   the    {\it
  GALEX}\footnote{http://galex.stsci.edu/GR2/?page=mastform}
\citep{Martin:2003qy}  source catalog  and  independent photometry  on
archival    images    provides    magnitudes   $FUV=23.4\pm0.1$    and
$NUV=19.04\pm0.02$.

A likely explanation for FE-1 is an eclipsing contact binary system of
W~Uma type with an  orbital period of $0.38050\pm0.00004$\,days.  This
is supported by the apparent  two minima in the lightcurve which could
indicate the phase of primary and secondary occultation.

\begin{figure}[htbp]
\begin{center}
\includegraphics[width=0.35\textwidth,angle=90]{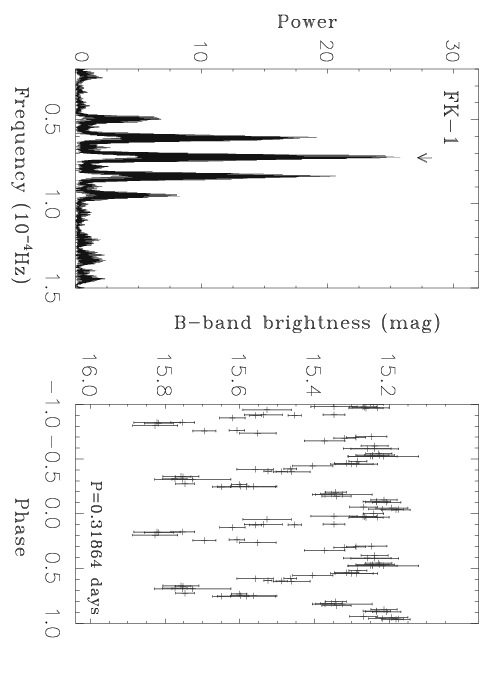}
\caption{Same as Figure~\ref{fig:phaseE} for FK-1. The highest power occurs at a frequency of $(7.264\pm0.002)\times10^{-5}$\,Hz (or a period of $0.15932\pm0.00003$\,days). The phase  folded  heliocentric  corrected lightcurve again indicates differences in the depth of consecutive minima, suggesting that the real period is likely $0.31864\pm0.00006$\,days.}
\label{fig:phaseK}
\end{center}
\end{figure}

The  second   periodic  variable,  FK-1   (Figure~\ref{fig:fc}e),  was
observed  as a  point source  with  mean brightness  of $B=15.42$  and
regular    non-sinusoidal     variations    with    a     period    of
$0.31864\pm0.00006$\,days     (Figures~\ref{fig:lc}e,\ref{fig:phaseK}).
The  maximum photometric  amplitude  was $\Delta  B=0.65\pm0.07$\,mag.
The   source  has  a   bright  2MASS   counterpart  ($J=13.42\pm0.03$,
$K=12.93\pm0.03$)    and     its    colors    ($B-J=2.00\pm0.20$\,mag,
$B-K=2.49\pm0.20$\,mag) resemble those of a G0-9 giant or dwarf.

While the short modulation  and lightcurve shape in principle resemble
also  that of  an RRc~Lyrae  pulsator, the  spectral type  of  G5-9 is
inconsistent with this  classification. The most likely interpretation
for FK-1 is, similar to FE-1, a W~Uma type eclipsing binary.

\begin{figure}[htbp]
\begin{center}
\includegraphics[width=0.35\textwidth,angle=90]{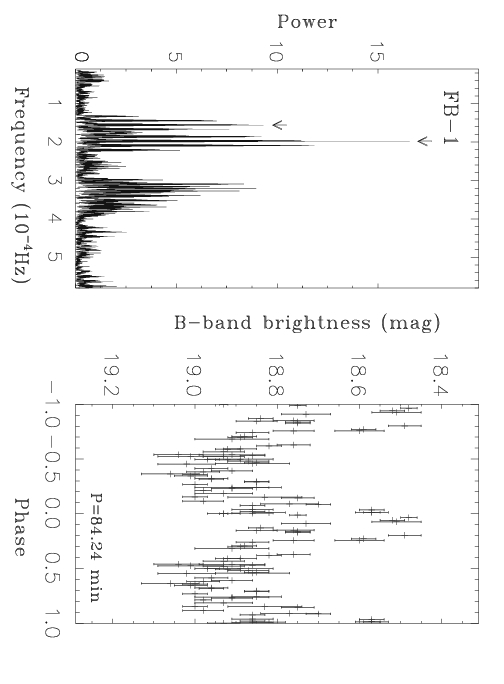}
\caption{Same  as Figure~\ref{fig:phaseE}  for  FB-1. The  periodogram
  shows  a  large  number  of   peaks,  with  the  most  prominent  at
  $(1.975\pm0.005)\times10^{-4}$\,Hz ($84.24\pm0.15$\,min). A second period occurs at ~107\,min.}
\label{fig:phaseB}
\end{center}
\end{figure}

The  classification   of  the   remaining  variable,  FB-1,   is  more
challenging. Its  lightcurve displayed apparently  erratic variability
with   an   rms  of   0.14\,mag   around   the   mean  brightness   of
$B=18.84\pm0.14$   and   with   a   maximum   amplitude   of   $\Delta
B=0.57\pm0.08$\,mag     (Figure~\ref{fig:lc}b).      A    Lomb-Scargle
periodogram (Figure~\ref{fig:phaseB}) revealed  a number of peaks, the
most   prominent  being   at  a   period   of  $0.0585\pm0.0001$\,days
($84.24\pm0.15$\,min).  However,  the folded lightcurve  shows several
data points that  deviate notably from this phasing.  We also not that
the  power  spectrum  shows  evidence  for  an  additional  period  of
$\sim107$\,min. Thus, the tentative  timescale of modulation has to be
taken  with  caution  and  confirmation by  observations  with  higher
sampling rate will be required.

\begin{figure}[htbp]
\begin{center}
\includegraphics[width=0.45\textwidth,angle=0]{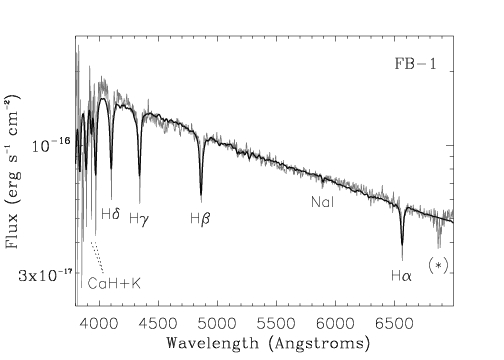}
\caption{NTT/EMMI spectrum  (grey) of FB-1 obtained on  2007 Dec 12.19
  UT using the low-dispersion  mode (RILD) with grism\#5 (instrumental
  FWHM=4.5\AA).  The  data were reduced with  customized IRAF routines
  and  flux  calibrated   in  comparison  to  the  spectro-photometric
  standard  star  LTT377 \citep{Hamuy:1992zr}.  The  black solid  line
  shows   an  A7V   template   spectrum  from  Pickles (1998).
  Prominent spectral features are indicated.}
\label{fig:spec}
\end{center}
\end{figure}

FB-1 has no 2MASS counterpart  to a limiting magnitude of $J=17$ which
constrains its  color to $B-J<1.84$\,mag.  This, together  with a {\it
  GALEX} detection at  NUV=21.39$^{+0.56}_{-0.37}$ and a non-detection
in the FUV filter to  $>21$\,mag restricts a stellar classification to
A-G.  A spectrum obtained with the ESO Multi-Mode Instrument (EMMI) at
the  NTT, La  Silla,  shows  a blue  continuum  with prominent  Balmer
absorption lines  (Figure~\ref{fig:spec}). A comparison  with template
stellar     spectra    \citep{Pickles:1998xy}    suggests     an    A7
classification. This spectral range  is populated by pulsators such as
$\delta$~Scuti  stars  and RR~Lyrae.   Of  those, only  $\delta$~Scuti
stars,  and the phenomenologically\footnote{$\delta$~Scuti  and SX~Phe
  stars   are  generally  distinguished   by  their   hosting  stellar
  population and  metallicity. This can  not be accomplished  with the
  available  data. }   similar SX~Phe  stars \citep{Rodriguez:1990uq},
exhibit modulations on timescales  as short as detected for FB-1.  As
indicated by the  number of short period peaks  in the periodogram, we
may have  to consider also  events with periods below  84\,mins.  Some
white  dwarf  systems  do  show  periodic  pulsations  below  30\,min.
ZZ~Ceti    type     sources    have    white     dwarf    temperatures
\citep[11--12\,kK;][]{Bergeron:2004lr} in  agreement with the observed
colors \citep{Kawka:2006uq}.  However, the lack of pressure broadening
of the  Balmer absorption  lines rules out  the white  dwarf scenario.
Awaiting  further test  of the  periodic  modulation, we  find that  a
Galactic  $\delta$~Scuti  or  SX~Phe  star  appears  to  be  the  most
intriguing interpretation for FB-1.

\section{Discussion}

Here were reported the results  of a dedicated search for optical fast
transients  and variables in  the environment  of the  Fornax~I galaxy
cluster.   In  total,  the  survey  netted  two  transients  and  five
high-amplitude   ($\Delta   B>0.1$\,mag)   variable   sources.    Both
transients have been detected as  a result of the inter-month sampling
of  our   observations  and  have  independently   been  reported  and
classified     as      type     Ia     supernovae      in     NGC~1216
\citep{Monard:2006lr,Monard:2006fk}.   The   properties  of  the  five
variables  suggests that  they are  located in  the foreground  to the
Fornax~I cluster and belong to  the Milky Way stellar population. This
is in agreement with the large projected distance (26--42\,kpc) of the
events      to     the      nearest      Fornax     galaxies      (see
Table~\ref{tab:candidates}).  Two of  the variables were identified as
flares from M-dwarfs, while other two are likely eclipsing binaries of
the W~Uma type.   The remaining variable shows indications  for a very
short periodicity ($\sim84$\,min), and  spectroscopy suggests it to be
a $\delta$\,Scuti or SX~Phe  star.  No genuine fast transient brighter
than $B=21.3$ has been detected.

Surveys, like the one presented  here, can offer valuable guidance for
the  planning  and  execution  of  the next  generation  of  transient
experiments.   In order to  provide relevant  event rates,  a complete
identification of all sources detected with a given choice of cadence,
filter, depth, amplitude, and  field selection, is required. This will
pose  inevitable  challenges   for  follow-up  limited  projects  like
Pan-STARRS  and  LSST. However,  at  least  for variables,  multi-band
photometry and good temporal coverage can already be strong indicators
for  the  nature   of  an  event,  as  demonstrated   for  the  Fornax
sources. Indeed,  automatic cross-matching  with UV to  near-IR source
catalogs with  a similar depth  as the surveys  will be vital  for the
success of the future large optical projects.

\subsection{All-Sky Rate of Fast Transients}

In the following,  we will discuss the estimates  and expectations for
transients  on  timescales of  $\sim32$\,min  and  with apparent  peak
brightness of $15<B<21.3$.  We start by determining the upper limit on
the rate  of cosmological fast  transients not associated with  any of
the Fornax  cluster galaxies.  As described  in \S~\ref{sec:data}, the
areal survey exposure is  $E_A=1.86$\,deg$^2$ and the efficiency for a
single  detection  of  a   field  source  brighter  than  $B=21.3$  is
$\epsilon=0.7$.  As we require a candidate to be found in at least two
images,  $\epsilon$ needs to  be squared.   The non-detection  of fast
transients in the survey translates  into a 95\,\% Poisson upper limit
of  N=$3$\,  events  \citep{Gehrels:1986qy}.  The  corresponding  rate
follows from

\begin{equation}
r_A=\frac{N}{\epsilon^2\cdot E_A}
\end{equation}

as  ${\rm r_A}  <3.3$\,  events day$^{-1}$  deg$^{-2}$,  or an  annual
all-sky rate of ${\rm  r_A}<5\times10^7$.  Note, that these limits are
valid only  for an event population that  is homogeneously distributed
over the  sky.  In  other words, the  detection probability has  to be
independent  of the field  selection.  Under  this assumption,  we can
match our  estimates with previously obtained results,  e.g., from the
Deep  Lens   Survey  transient  search  \citep[DLS;][]{Becker:2004fk}.
Indeed, the Fornax  survey is very similar to the  DLS with respect to
exposure  (1.38 vs  1.1  deg$^2$  days) and  probed  timescale (32  vs
22\,min).  As also no genuine fast transient was found in the DLS, the
the event rate is comparable.

\subsection{Rates of Fast Transients in Nearby Galaxies}

The main motivation of our survey was the exploration of the transient
and variable population in a  specific environment, namely in a galaxy
cluster with known distance and  known stellar mass. For this purpose,
we will  now calculate the specific  rate of fast  transients per unit
stellar mass, or its proxy, the $B$-band luminosity.

As discussed above (see also Fig.~\ref{fig:efficiency}), $\epsilon$ is
not  only a  function  of the  magnitude  of a  transient, $M_T$,  but
depends  also  on the  underlying  surface  brightness, m$_{\rm  SB}$.
Similarly, the product of  the luminosity, L$_{\rm SB,i}$, enclosed in
a surface  brightness bin, m$_{\rm SB,i}$, and  the relative exposure,
$t$, is different for all targeted Fornax galaxies.  Here, $t$ denotes
the time a given galaxy was probed for transients on timescales of the
32\,min  cadence. We  start  by calculating  L$_{\rm  SB,i}$ for  each
galaxy  listed in  Table~\ref{tab:fields}, with  $i$ running  from the
saturation  limit  of  16.7\,mag  asec$^{-2}$ to  the  sky  background
brightness of  22.6\,mag asec$^{-2}$ in  steps of 0.1\,mag.   Next, we
estimate the  survey exposure, $E_{SB,i}$, for each  m$_{\rm SB,i}$ in
units\footnote{Here, we  use an  absolute solar $B$-band  magnitude of
  M$_{B,\odot}=5.48$}  of  M$_{B,\odot}$   years  by  summing  L$_{\rm
  SB,i}\cdot t$ for  all galaxies. The rate of  transients as function
of the brightness can then be calculated as

\begin{equation}
r(M_T)=\sum_i{\frac{N(M_T)}{\epsilon(M_T,m_{\rm SB})_i^2\cdot E_{{\rm SB},i}}}
\end{equation}

Formerly, the number of detected  transients, $N$, also depends on the
peak brightness. However,  here we use again the  95\,\% Poisson upper
limit of $N(M_T)=N=3$  events, as no event in  the entire probed range
of   $15<B<21.3$   was   detected.  

\begin{figure}[htbp]
\begin{center}
\includegraphics[width=0.45\textwidth,angle=0]{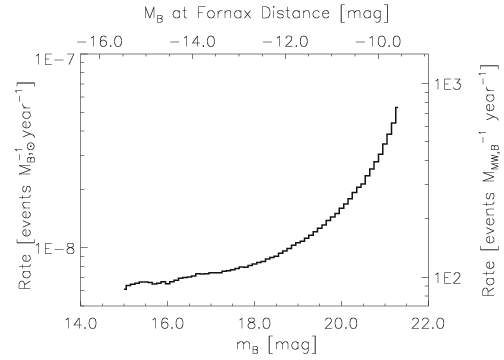}
\caption{95\,\% Poission upper limit on the rate of transients on timescales of 32\,min as function of peak brightness.}
\label{fig:rates}
\end{center}
\end{figure}

The result  is presented in Figure~\ref{fig:rates}, where  we show the
upper  limit  on the  rate  as  funtion  of transient  magnitude.   As
expected,  the  brightest events  (m$_B<$18.5,  M$_B<-12.4$) have  the
lowest limits  ($r<10^{-8}$ events M$_{B,\odot}^{-1}$  yr$^{-1}$). For
the   Milky   Way   disk   \citep[$M_B=-19.9$;][]{Quillen:1997}   this
translates into  $<140$ events per year,  e.g., a few  times more than
the classical novae rate.  One could argue that any Galactic transient
population  brighter  than  classical  novae  and  with  a  comparable
frequency  may  have  been  discovered already.   However,  the  short
timescales tested in our survey require a significantly higher cadence
than typically  used in transient searches.  Fast  transients may have
previously  been detected in  single images,  but simply  discarded as
unconfirmed or spurious  events. However, there have been  a number of
planetary transit and variable  star studies, which offer the required
high cadence \citep[e.g.,;][]{Kane:2005xx,Kraus:2007xx,Lister:2007xx}.
These experiments  typically target dense  Galactic fields and  thus a
large  stellar population.  Unfortunately,  results on  transients, as
opposed to transits, are rarely recorded or published.

\section{Conclusion}

The  existence  of fast  transients  requires  further constraint  with
specialized experiments.   The most exciting facility  in this respect
will be the  Palomar Transient Factory (PTF, Rau et  al., in prep.), a
dedicated transient survey instrument at the Palomar 48-inch telescope
with expected  first light  in Winter  2008.  The PTF  will use  a 7.8
square degree field of view  to perform a number of transient searches
on various time scales. The large field of view will allow us to cover
galaxy  clusters  with  fewer  pointings, thus  increasing  the  total
stellar mass observed at a given cadence.

However, we have  shown that the probability of  detecting a transient
strongly depends on the local  surface brightness. While this is not a
surprising  result,  its  relevance  for  future  searches  in  nearby
galaxies needs to  be understood. There will always be  the need for a
compromise between limiting  magnitude of an event to  be detected and
galaxy fraction covered.  Larger apertures or longer exposures are not
necessarily   improving   the    detection   probability   for   faint
events. Although more source photons will be collected, also more host
light is recieved and and a  larger extend of the galaxy will be above
saturation. The main factor for  success will be the image quality. At
a  given  surface  brightness,   a  narrower  PSF  will  increase  the
signal-to-noise, and thus, the probability for a detection.


\acknowledgments 

We are grateful to Paul Price for providing his image subtraction code
and to  Carrol Wainwright  for his contribution  to the  pipeline.  We
thank  Mara  Salvato,  Mansi  M.~Kasliwal  and  S.~Bradley  Cenko  for
discussion and  constructive criticism. We  thank the referee  for his
stimulating comments and suggestions. This work is based on in part on
observations  collected at the  European Southern  Observatory, Chile.
This publication  makes use of data  products from the  Two Micron All
Sky  Survey,   which  is  a   joint  project  of  the   University  of
Massachusetts    and   the    Infrared    Processing   and    Analysis
Center/California  Institute  of Technology,  funded  by the  National
Aeronautics  and   Space  Administration  and   the  National  Science
Foundation. This research has  made use of the NASA/IPCA Extragalactic
Database  (NED) which is  operated by  the Jet  Propulsion Laboratory,
California Institute  of Technology, under contract  with the National
Aeronautics and  Space Administration. This work is  supported in part
by grants from the National Science Foundation and NASA.




\bibliographystyle{apj}

\begin{thebibliography}{}
\bibitem[{Alard} 2000]{Alard:2000fk}
{Alard}, C. 2000, \aaps, 144, 363

\bibitem[{Becker} et al. 2004]{Becker:2004fk}
{Becker}, A.~C., {Wittman}, D.~M., {Boeshaar}, P.~C., {et al.} 2004, \apj, 611, 
418

\bibitem[{Bergeron} et al. 2004]{Bergeron:2004lr}
{Bergeron}, P., {Fontaine}, G., {Bill{\`e}res}, M., {et al.} 2004, \apj, 600, 404

\bibitem[{Bertin} \& {Arnouts} 1996]{Bertin:1996uq}
{Bertin}, E. \& {Arnouts}, S. 1996, \aaps, 117, 393

\bibitem[{Gehrels} 1986]{Gehrels:1986qy}
{Gehrels}, N. 1986, \apj, 303, 336

\bibitem[{Grillmair} et al. 1999]{Grillmair:1999lr}
{Grillmair}, C.~J., {Forbes}, D.~A., {Brodie}, J.~P., {et al.} 1999, \aj, 117, 167

\bibitem[{Hamuy} et al. 1992]{Hamuy:1992zr}
{Hamuy}, M., {Walker}, A.~R., {Suntzeff}, N.~B., {et al.} 1992, \pasp, 104, 533

\bibitem[{Kaiser} et al. 2002]{Kaiser:2002fk}
{Kaiser}, N., {Aussel}, H., {Burke}, B.~E., {et al.} 2002, in Presented at the Society of  Photo-Optical Instrumentation Engineers (SPIE) Conference, Vol. 4836, Survey  and Other Telescope Technologies and Discoveries. Edited by Tyson, J.  Anthony; Wolff, Sidney. Proceedings of the SPIE, Volume 4836, pp. 154-164  (2002)., ed. J.~A. {Tyson} \& S.~{Wolff}, 154

\bibitem[{Kane} et al. 2005]{Kane:2005xx}
Kane, S.~R., Lister, T.~A., Cameron, A.~C., et al. 2005, \mnras, 362, 117

\bibitem[{Kasliwal} et al. 2007]{Kasliwal:2007qy}
{Kasliwal}, M.~M., {Cenko}, S.~B., {Kulkarni}, S.~R., {et al.} 2007, ArXiv e-prints, 0708.0226

\bibitem[{Kawka} et al. 2006]{Kawka:2006uq}
{Kawka}, A., {Vennes}, S., {Oswalt}, T.~D., {et al.} 2006, \apjl, 643, L123

\bibitem[{Klotz} et al. 2007]{Klotz:2007lr}
{Klotz}, A., {Boer}, M., \& {Atteia}, J. 2007, GRB Coordinates Network, 6769, 1

\bibitem[{Kraus} et al. 2007]{Kraus:2007xx}
{Kraus}, A.~L., Craine, E.~R., Giampapa, M.~S., et al. 2007, AJ, 134, 1488

\bibitem[{Kulkarni} et al. 2007]{Kulkarni:2007uq}
{Kulkarni}, S.~R., {Ofek}, E.~O., {Rau}, A., {et al.} 2007,  \nat, 447, 458

\bibitem[{Kulkarni} \& {Rau} 2006]{Kulkarni:2006lr}
{Kulkarni}, S.~R. \& {Rau}, A. 2006, \apjl, 644, L63

\bibitem[{Landolt} 1992]{Landolt:1992lr}
{Landolt}, A.~U. 1992, \aj, 104, 340

\bibitem[{Lister} et al. 2007]{Lister:2007xx}
Lister, T.~A., West, R.~G., Wilson, D.~M., et al. 2007, \mnras, 379, 647

\bibitem[{Martin} et al. 2003]{Martin:2003qy}
{Martin}, C., {Barlow}, T., {Barnhart}, W., {et al.} 2003, in Presented at the Society of Photo-Optical  Instrumentation Engineers (SPIE) Conference, Vol. 4854, Future EUV/UV and  Visible Space Astrophysics Missions and Instrumentation. Edited by J. Chris  Blades, Oswald H. W. Siegmund. Proceedings of the SPIE, Volume 4854, pp.  336-350 (2003)., ed. J.~C. {Blades} \& O.~H.~W. {Siegmund}, 336

\bibitem[{Monard} 2006]{Monard:2006lr}
{Monard}, L.~A.~G. 2006, Central Bureau Electronic Telegrams, 553, 1

\bibitem[{Monard} \& {Folatelli} 2006]{Monard:2006fk}
{Monard}, L.~A.~G. \& {Folatelli}, G. 2006, Central Bureau Electronic  Telegrams, 723, 1

\bibitem[{Morales-Rueda} et al. 2006]{Morales-Rueda:2006qy}
{Morales-Rueda}, L., {Groot}, P.~J., {Augusteijn}, T., {et al.} 2006, \mnras, 371, 1681

\bibitem[{Ofek} et al. 2007a]{Ofek:2007fj}
{Ofek}, E.~O., {Cameron}, P.~B., {Kasliwal}, M.~M., {et al.} 2007a, \apjl, 659, L13

\bibitem[{Ofek} et al. 2007b]{Ofek:2007uq}
{Ofek}, E.~O., {Kulkarni}, S.~R., {Rau}, A., {et al.} 2007b, ArXiv e-prints, 710

\bibitem[{Pagani} et al. 2007]{Pagani:2007fk}
{Pagani}, C., {Barthelmy}, S.~D., {Cummings}, J.~R., {et al.} 2007, GRB Coordinates Network, 6489, 1

\bibitem[{Pastorello} et al. 2007]{Pastorello:2007jk}
{Pastorello}, A., {Smartt}, S.~J., {Mattila}, S., {et al.} 2007, Nature, 447, 829

\bibitem[{Pickles} 1998]{Pickles:1998xy}
{Pickles}, A.~J. 1998, \pasp, 110, 863

\bibitem[{Quillen} \& {Sarajedini} 1997]{Quillen:1997}
Quillen, A.~C., \& Sarajedini, V.~L., 1997, AJ, 115, 1412 

\bibitem[{Quimby} et al. 2007]{Quimby:2007qy}
{Quimby}, R.~M., {Aldering}, G., {Wheeler}, J.~C., {et al.} 2007, \apjl, 668, L99

\bibitem[{Ramsay} et al. 2006]{Ramsay:2006lr}
{Ramsay}, G., {Napiwotzki}, R., {Hakala}, P., {et al.} 2006, \mnras, 371,  957

\bibitem[{Rau} et al. 2007]{Rau:2007kx}
{Rau}, A., {Kulkarni}, S.~R., {Ofek}, E.~O., {et al.} 2007, \apj, 659, 1536

\bibitem[{Rodriguez} et al. 1990]{Rodriguez:1990uq}
{Rodriguez}, E., {Rolland}, A., \& {Lopez de Coca}, P. 1990, \apss, 169, 113

\bibitem[{Rykoff} et al. 2005]{Rykoff:2005lr}
{Rykoff}, E.~S., {Aharonian}, F., {Akerlof}, C.~W., {et al.} 2005, \apj, 631, 1032

\bibitem[{Scargle} 1982]{Scargle:1982kx}
{Scargle}, J.~D. 1982, \apj, 263, 835

\bibitem[{Schlegel} et al. 1998]{Schlegel:1998ul}
{Schlegel}, D.~J., {Finkbeiner}, D.~P., {Davis}, M., 1998, ApJ, 500, 525

\bibitem[{Schmidt} et al. 2005]{Schmidt:2005qy}
{Schmidt}, B.~P., {Keller}, S.~C., {Francis}, P.~J., {et al.} 2005,  in Bulletin of the American Astronomical Society, Vol.~37, Bulletin of the  American Astronomical Society, 457

\bibitem[{Smith} et al  2007]{Smith:2007qy}
{Smith}, N., {Li}, W., {Foley}, R.~J., {et al.} 2007, \apj, 666, 1116

\bibitem[{Stefanescu} et al. 2007]{Stefanescu:2007fk}
{Stefanescu}, A., {Slowikowska}, A., {Kanbach}, G., {et al.} 2007, GRB Coordinates Network, 6508, 1

\bibitem[{Tyson} 2005]{Tyson:2005lr}
{Tyson}, A. 2005, in Astronomical Society of the Pacific Conference Series,  Vol. 339, Observing Dark Energy, ed. S.~C. {Wolff} \& T.~R. {Lauer}, 95

\bibitem[{van Dokkum} 2001]{van-Dokkum:2001qy}
{van Dokkum}, P.~G. 2001, \pasp, 113, 1420

\bibitem[{Voges} et al. 1999]{Voges:1999fj}
{Voges}, W., {Aschenbach}, B., {Boller}, T., {et al.} 1999, \aap, 349, 389

\bibitem[{Zacharias} et al. 2005]{Zacharias:2005lr}
{Zacharias}, N., {Monet}, D.~G., {Levine}, S.~E., {et al.} 2005, VizieR Online Data Catalog, 1297, 0
\end{thebibliography}

\LongTables

\tabletypesize{\normalsize}
\def\arraystretch{5.3}
\begin{longtable}{ccc}
\hline \hline
MJD & B-band Mag & Err \\
\hline \hline
\endfirsthead
\hline\hline
 MJD & B-band Mag & Err \\
\hline\hline
\endhead
\multicolumn{3}{c}{\bf FA-1}\\
54031.09131  &  19.14 & 0.07 \\
54031.13743  &  19.18 & 0.07 \\
54031.16762  &  19.14 & 0.06 \\
54031.19747  &  19.19 & 0.06 \\
54031.22753  &  19.17 & 0.05 \\ 
54031.25564  &  19.17 & 0.12 \\
54031.28432  &  19.23 & 0.04 \\
54031.31550  &  19.20 & 0.03 \\
54031.34153  &  19.17 & 0.04 \\
54031.36836  &  19.16 & 0.05 \\
54032.09634  &  19.13 & 0.05 \\
54032.11925  &  19.18 & 0.07 \\
54032.14265  &  19.14 & 0.07 \\
54032.16545  &  19.14 & 0.05 \\
54032.18746  &  19.17 & 0.05 \\
54032.20956  &  19.19 & 0.04 \\
54032.23146  &  19.17 & 0.03 \\
54032.25373  &  19.20 & 0.04 \\
54032.27616  &  19.18 & 0.04 \\
54032.29918  &  19.16 & 0.04 \\
54032.32214  &  19.13 & 0.04 \\
54032.34450  &  19.16 & 0.05 \\
54032.36704  &  19.16 & 0.07 \\
54056.25381  &  19.21 & 0.06 \\
54057.13576  &  19.20 & 0.04 \\
54058.17187  &  19.14 & 0.04 \\
54059.21876  &  19.20 & 0.03 \\
54060.21648  &  19.16 & 0.04 \\
54088.03730  &  19.20 & 0.05 \\
54088.06037  &  19.19 & 0.05 \\
54088.08310  &  18.91 & 0.03 \\
54088.10545  &  19.08 & 0.05 \\
54088.12831  &  19.11 & 0.04 \\
54088.15113  &  19.11 & 0.05 \\
54088.17361  &  19.12 & 0.04 \\
54088.19600  &  19.17 & 0.05 \\
54088.21828  &  19.14 & 0.05 \\
54088.24087  &  19.14 & 0.04 \\
54088.26362  &  19.17 & 0.04 \\
54088.28586  &  19.19 & 0.08 \\
54088.30859  &  19.12 & 0.04 \\
54088.33179  &  19.14 & 0.08 \\
54089.04811  &  19.17 & 0.05 \\
54089.07056  &  19.17 & 0.04 \\
54089.09291  &  19.15 & 0.05 \\
54089.11489  &  19.16 & 0.06 \\
54089.13760  &  19.19 & 0.04 \\
54089.15988  &  19.13 & 0.04 \\
54089.18222  &  19.10 & 0.03 \\
54089.20416  &  19.14 & 0.05 \\
54089.22648  &  19.17 & 0.05 \\
54089.24890  &  19.12 & 0.04 \\
54089.27174  &  19.14 & 0.04 \\
54089.29370  &  19.14 & 0.04 \\
54089.31590  &  19.13 & 0.06 \\
54090.07199  &  19.21 & 0.04 \\
54090.09391  &  19.18 & 0.06 \\
54090.11576  &  19.18 & 0.04 \\
54090.13819  &  19.21 & 0.04 \\
54090.16077  &  19.17 & 0.04 \\
54090.18262  &  19.19 & 0.06 \\
54090.20461  &  19.23 & 0.06 \\
54090.22677  &  19.20 & 0.06 \\
54090.24872  &  19.18 & 0.05 \\
54090.27118  &  19.20 & 0.06 \\
54090.29136  &  19.22 & 0.09 \\
\hline                       
\multicolumn{3}{c}{\bf FB-1}\\
54056.24324  &  18.93 & 0.01 \\
54031.08625  &  19.01 & 0.04 \\
54031.13400  &  18.86 & 0.03 \\
54031.16505  &  18.93 & 0.04 \\
54031.19449  &  18.89 & 0.08 \\
54031.22439  &  18.75 & 0.02 \\
54031.25312  &  18.97 & 0.03 \\
54031.28150  &  18.57 & 0.02 \\
54031.33885  &  18.57 & 0.03 \\
54031.36487  &  18.92 & 0.03 \\
54032.07101  &  19.04 & 0.06 \\
54032.09415  &  18.75 & 0.04 \\
54032.11686  &  18.76 & 0.05 \\
54032.13991  &  19.06 & 0.07 \\
54032.16328  &  18.48 & 0.02 \\
54032.18511  &  18.89 & 0.04 \\
54032.20738  &  18.98 & 0.02 \\
54032.22929  &  18.75 & 0.03 \\
54032.25156  &  18.86 & 0.09 \\
54032.27399  &  18.86 & 0.03 \\
54032.29702  &  18.91 & 0.09 \\
54032.31996  &  18.85 & 0.03 \\
54032.34229  &  18.73 & 0.06 \\
54032.36487  &  19.01 & 0.08 \\
54056.24324  &  18.86 & 0.03 \\
54057.13337  &  18.75 & 0.03 \\
54058.16927  &  18.83 & 0.09 \\
54059.21565  &  18.85 & 0.03 \\
54060.21333  &  19.00 & 0.03 \\
54088.03266  &  18.93 & 0.05 \\
54088.03495  &  18.87 & 0.05 \\
54089.04593  &  18.91 & 0.05 \\
54088.05818  &  18.70 & 0.03 \\
54088.08092  &  18.89 & 0.03 \\
54088.10322  &  18.96 & 0.03 \\
54088.12609  &  18.51 & 0.06 \\
54088.14885  &  18.91 & 0.08 \\
54088.17140  &  19.00 & 0.03 \\
54088.19379  &  18.59 & 0.03 \\
54088.21604  &  18.91 & 0.05 \\
54088.23868  &  18.86 & 0.05 \\
54088.26127  &  18.82 & 0.03 \\
54088.28364  &  18.91 & 0.09 \\
54088.30637  &  18.76 & 0.05 \\
54088.32957  &  19.02 & 0.07 \\
54089.06839  &  18.85 & 0.05 \\
54089.09070  &  18.85 & 0.03 \\
54089.11270  &  18.77 & 0.07 \\
54089.13518  &  18.88 & 0.03 \\
54089.15769  &  18.96 & 0.04 \\
54089.18005  &  18.52 & 0.03 \\
54089.20199  &  18.93 & 0.05 \\
54089.22429  &  18.93 & 0.07 \\
54089.24672  &  18.49 & 0.04 \\
54089.26953  &  18.96 & 0.04 \\
54089.29153  &  18.86 & 0.05 \\
54090.06978  &  18.60 & 0.06 \\
54090.09173  &  18.98 & 0.04 \\
54090.11358  &  18.82 & 0.04 \\
54090.13593  &  18.76 & 0.04 \\
54090.15861  &  18.85 & 0.06 \\
54090.18034  &  18.84 & 0.03 \\
54090.20245  &  18.92 & 0.03 \\
54090.22459  &  18.98 & 0.05 \\
54090.24656  &  18.86 & 0.04 \\
54090.26901  &  19.00 & 0.03 \\
54090.28920  &  18.57 & 0.04 \\
54090.31140  &  18.85 & 0.05 \\
\hline
\multicolumn{3}{c}{\bf FE-1}\\   
54031.10145  &  16.27 & 0.01 \\
54031.14394  &  16.09 & 0.03 \\
54031.17524  &  16.07 & 0.01 \\
54031.20451  &  16.12 & 0.03 \\
54031.23271  &  16.22 & 0.01 \\
54031.26175  &  16.28 & 0.03 \\
54031.29416  &  16.23 & 0.04 \\
54031.32096  &  16.12 & 0.03 \\
54031.34712  &  16.08 & 0.04 \\
54031.37445  &  16.09 & 0.02 \\
54032.10104  &  16.10 & 0.01 \\
54032.12414  &  16.07 & 0.01 \\
54032.14715  &  16.10 & 0.03 \\
54032.16982  &  16.18 & 0.02 \\
54032.19197  &  16.32 & 0.01 \\
54032.21402  &  16.36 & 0.01 \\
54032.23611  &  16.32 & 0.01 \\
54032.25832  &  16.18 & 0.01 \\
54032.28061  &  16.09 & 0.02 \\
54032.30355  &  16.07 & 0.02 \\
54032.32675  &  16.08 & 0.03 \\
54032.34903  &  16.13 & 0.02 \\
54032.37144  &  16.21 & 0.02 \\
54056.26076  &  16.08 & 0.02 \\
54057.14122  &  16.29 & 0.03 \\
54058.17666  &  16.06 & 0.01 \\
54059.22368  &  16.33 & 0.01 \\
54060.22240  &  16.18 & 0.01 \\
54088.04208  &  16.07 & 0.01 \\
54088.06537  &  16.07 & 0.01 \\
54088.08778  &  16.12 & 0.01 \\
54088.11031  &  16.22 & 0.01 \\
54088.13270  &  16.35 & 0.01 \\
54088.15555  &  16.34 & 0.01 \\
54088.17821  &  16.24 & 0.02 \\
54088.20043  &  16.14 & 0.04 \\
54088.22283  &  16.09 & 0.05 \\
54088.24545  &  16.08 & 0.03 \\
54088.26806  &  16.12 & 0.05 \\
54088.29067  &  16.18 & 0.05 \\
54088.31310  &  16.29 & 0.03 \\
54088.33641  &  16.30 & 0.04 \\
54089.05255  &  16.18 & 0.02 \\
54089.07493  &  16.28 & 0.04 \\
54089.09732  &  16.29 & 0.03 \\
54089.11962  &  16.29 & 0.03 \\
54089.14196  &  16.17 & 0.01 \\
54089.16446  &  16.12 & 0.03 \\
54089.18674  &  16.08 & 0.04 \\
54089.20877  &  16.09 & 0.04 \\
54089.23087  &  16.14 & 0.02 \\
54089.25336  &  16.25 & 0.05 \\
54089.27618  &  16.35 & 0.03 \\
54089.29822  &  16.35 & 0.05 \\
54089.32037  &  16.26 & 0.05 \\
54090.07644  &  16.27 & 0.03 \\
54090.09791  &  16.15 & 0.02 \\
54090.12009  &  16.08 & 0.04 \\
54090.14271  &  16.07 & 0.01 \\
54090.16510  &  16.09 & 0.05 \\
54090.18709  &  16.15 & 0.02 \\
54090.20928  &  16.25 & 0.02 \\
54090.23116  &  16.31 & 0.04 \\
54090.25306  &  16.30 & 0.01 \\
54090.27555  &  16.22 & 0.06 \\
54090.29584  &  16.14 & 0.03 \\
54090.31800  &  16.12 & 0.05 \\
54031.10929  &  16.22 & 0.03 \\
54031.14870  &  16.07 & 0.03 \\
54031.17989  &  16.07 & 0.01 \\
54031.21009  &  16.13 & 0.01 \\
54031.23783  &  16.25 & 0.02 \\
54031.26771  &  16.29 & 0.05 \\
54031.29886  &  16.22 & 0.03 \\
54031.32636  &  16.13 & 0.01 \\
54031.35253  &  16.08 & 0.01 \\
54032.15194  &  16.11 & 0.01 \\
54032.17424  &  16.20 & 0.01 \\
54032.19631  &  16.34 & 0.01 \\
54032.21836  &  16.36 & 0.01 \\
54032.24047  &  16.29 & 0.01 \\
54032.26266  &  16.16 & 0.02 \\
54032.28498  &  16.10 & 0.01 \\
54032.30789  &  16.07 & 0.01 \\
54032.33108  &  16.10 & 0.01 \\
54032.35369  &  16.15 & 0.02 \\
54056.26546  &  16.08 & 0.02 \\
54057.14590  &  16.29 & 0.01 \\
54058.18190  &  16.06 & 0.01 \\
54060.22853  &  16.17 & 0.02 \\
54088.04652  &  16.07 & 0.01 \\
54088.06976  &  16.08 & 0.01 \\
54088.09222  &  16.14 & 0.05 \\
54088.11479  &  16.26 & 0.01 \\
54088.13731  &  16.36 & 0.01 \\
54088.16018  &  16.35 & 0.02 \\
54088.18261  &  16.22 & 0.03 \\
54088.20486  &  16.14 & 0.02 \\
54088.22732  &  16.08 & 0.01 \\
54088.25000  &  16.09 & 0.01 \\
54088.27247  &  16.14 & 0.01 \\
54088.29510  &  16.20 & 0.02 \\
54088.31750  &  16.32 & 0.04 \\
54088.34082  &  16.31 & 0.04 \\
54089.03496  &  16.12 & 0.02 \\
54089.05701  &  16.19 & 0.02 \\
54089.07930  &  16.28 & 0.03 \\
54089.10172  &  16.28 & 0.03 \\
54089.12406  &  16.25 & 0.03 \\
54089.14655  &  16.15 & 0.02 \\
54089.16883  &  16.10 & 0.05 \\
54089.19107  &  16.09 & 0.01 \\
54089.21313  &  16.12 & 0.01 \\
54089.23525  &  16.18 & 0.01 \\
54089.25774  &  16.27 & 0.03 \\
54089.28052  &  16.37 & 0.05 \\
54089.30256  &  16.36 & 0.05 \\
54089.32501  &  16.23 & 0.03 \\
54090.08082  &  16.25 & 0.01 \\
54090.10266  &  16.15 & 0.02 \\
54090.12502  &  16.09 & 0.02 \\
54090.14724  &  16.08 & 0.02 \\
54090.16944  &  16.11 & 0.02 \\
54090.19143  &  16.18 & 0.01 \\
54090.21362  &  16.29 & 0.04 \\ 
54090.23560  &  16.31 & 0.02 \\
54090.25750  &  16.30 & 0.04 \\
54090.27916  &  16.21 & 0.02 \\
54090.30019  &  16.14 & 0.02 \\
54090.32273  &  16.11 & 0.02 \\
\hline
\multicolumn{3}{c}{\bf FH-1}\\   
54031.12391  &  20.60 & 0.11 \\
54031.15980  &  20.80 & 0.13 \\
54031.18924  &  20.72 & 0.13 \\
54031.21800  &  20.77 & 0.14 \\
54031.24748  &  20.66 & 0.12 \\
54031.27642  &  20.73 & 0.13 \\
54031.30760  &  20.75 & 0.13 \\
54031.35971  &  20.67 & 0.13 \\
54032.11228  &  20.55 & 0.13 \\
54032.13553  &  20.80 & 0.14 \\
54032.15846  &  20.77 & 0.14 \\
54032.18078  &  20.81 & 0.14 \\
54032.20293  &  20.60 & 0.11 \\
54032.22490  &  20.80 & 0.14 \\
54032.24703  &  20.70 & 0.11 \\
54032.26939  &  20.75 & 0.12 \\
54032.29248  &  20.69 & 0.12 \\
54032.31453  &  20.78 & 0.13 \\
54032.33788  &  20.69 & 0.12 \\
54032.36031  &  20.74 & 0.14 \\
54088.05344  &  20.76 & 0.16 \\
54088.07640  &  20.77 & 0.16 \\
54088.09884  &  20.82 & 0.17 \\
54088.14412  &  20.74 & 0.15 \\
54088.16685  &  20.76 & 0.16 \\
54088.18928  &  20.65 & 0.14 \\
54088.21148  &  20.70 & 0.15 \\
54088.23406  &  20.60 & 0.15 \\
54088.25675  &  18.66 & 0.03 \\
54088.27912  &  19.78 & 0.07 \\
54088.30173  &  20.34 & 0.13 \\
54088.32432  &  20.68 & 0.18 \\
54088.34824  &  20.70 & 0.28 \\
54089.06390  &  20.64 & 0.11 \\ 
54089.08591  &  20.66 & 0.11 \\
54089.10831  &  20.68 & 0.12 \\
54089.13065  &  20.66 & 0.11 \\ 
54089.15326  &  20.62 & 0.12 \\
54089.19762  &  20.61 & 0.10 \\
54089.21987  &  20.60 & 0.11 \\ 
54089.24188  &  20.74 & 0.13 \\
54089.26443  &  20.62 & 0.12 \\
54089.28712  &  20.62 & 0.14 \\
54089.30938  &  20.70 & 0.15 \\
54089.33192  &  20.62 & 0.17 \\
54090.08737  &  20.88 & 0.14 \\
54090.10922  &  20.89 & 0.14 \\
54090.15416  &  20.66 & 0.13 \\
54090.17598  &  20.60 & 0.11 \\
54090.19805  &  20.75 & 0.14 \\
54090.24210  &  20.75 & 0.14 \\
54090.26442  &  20.63 & 0.14 \\
54090.30681  &  20.60 & 0.15 \\
54090.32935  &  20.80 & 0.28 \\
\hline
\multicolumn{3}{c}{\bf FK-1}\\    
54031.12012  &  15.24 & 0.02 \\
54031.15399  &  15.24 & 0.04 \\
54031.15669  &  15.25 & 0.03 \\
54031.18529  &  15.40 & 0.05 \\
54031.21552  &  15.75 & 0.04 \\
54031.24486  &  15.56 & 0.06 \\
54031.27346  &  15.34 & 0.03 \\
54031.30480  &  15.26 & 0.03 \\
54031.33144  &  15.34 & 0.06 \\
54031.35715  &  15.55 & 0.01 \\
54032.10997  &  15.22 & 0.04 \\
54032.13333  &  15.31 & 0.03 \\
54032.15628  &  15.52 & 0.01 \\
54032.17861  &  15.77 & 0.03 \\
54032.20066  &  15.59 & 0.02 \\
54032.22271  &  15.35 & 0.02 \\
54032.24487  &  15.23 & 0.03 \\
54032.26713  &  15.21 & 0.03 \\
54032.28934  &  15.26 & 0.03 \\
54032.31236  &  15.45 & 0.01 \\
54032.33545  &  15.81 & 0.03 \\
54032.35803  &  15.69 & 0.03 \\
54056.27284  &  15.24 & 0.04 \\
54057.15052  &  15.52 & 0.06 \\
54058.18653  &  15.28 & 0.02 \\
54059.23539  &  15.55 & 0.04 \\
54060.23351  &  15.74 & 0.02 \\
54088.05096  &  15.27 & 0.03 \\
54088.07419  &  15.53 & 0.05 \\
54088.09664  &  15.82 & 0.06 \\
54088.11924  &  15.60 & 0.02 \\
54088.14190  &  15.31 & 0.02 \\
54088.16463  &  15.23 & 0.03 \\
54088.18707  &  15.22 & 0.03 \\
54088.20927  &  15.28 & 0.02 \\
54088.23172  &  15.46 & 0.01 \\
54088.25450  &  15.75 & 0.02 \\
54088.27691  &  15.60 & 0.02 \\
54088.29950  &  15.34 & 0.02 \\
54088.32189  &  15.21 & 0.03 \\
54088.34545  &  15.19 & 0.04 \\
54089.03933  &  15.62 & 0.03 \\
54089.06167  &  15.82 & 0.06 \\
54089.08369  &  15.55 & 0.04 \\
54089.10613  &  15.37 & 0.05 \\
54089.12843  &  15.25 & 0.06 \\
54089.15093  &  15.24 & 0.06 \\
54089.17340  &  15.28 & 0.05 \\
54089.19542  &  15.48 & 0.01 \\
54089.21762  &  15.75 & 0.02 \\
54089.23961  &  15.58 & 0.07 \\
54089.26220  &  15.32 & 0.08 \\
54089.28494  &  15.20 & 0.03 \\
54089.30691  &  15.18 & 0.03 \\
54089.32940  &  15.26 & 0.04 \\
54090.08517  &  15.23 & 0.06 \\
54090.10701  &  15.21 & 0.09 \\
54090.12936  &  15.30 & 0.05 \\
54090.15200  &  15.46 & 0.04 \\
54090.17381  &  15.72 & 0.10 \\
54090.19578  &  15.64 & 0.02 \\
54090.21796  &  15.37 & 0.02 \\
54090.23994  &  15.23 & 0.03 \\
54090.26183  &  15.17 & 0.01 \\
54090.28424  &  15.23 & 0.03 \\
54090.30454  &  15.34 & 0.02 \\
54090.32707  &  15.75 & 0.03 \\
\hline
\caption{Data table for variables found in the survey. (for electronic version)}
\label{tab:data}
\end{longtable}



\clearpage

\end{document}